# New Directions in Science Emerge from Disconnection and Discord

Yiling Lin[1][2][3], James Evans[1][2]*, Lingfei Wu[3]*

Abstract

Science is built on scholarly consensus that shifts with time. This raises the question of how new and revolutionary ideas are evaluated and become accepted into the canon of science. Using two recently proposed metrics, we identify papers with high atypicality, which models how research draws upon novel combinations of prior research, and evaluate disruption, which captures the degree to which a study creates a new direction by eclipsing its intellectual forebears. Atypical papers are nearly two times more likely to disrupt science than conventional papers, but this is a slow process taking ten years or longer for disruption scores to converge. We provide the first computational model reformulating atypicality as the distance across latent knowledge spaces learned by neural networks. The evolution of this knowledge space characterizes how yesterday's novelty forms today's scientific conventions, which condition the novelty—and surprise—of tomorrow's breakthroughs.

*Keywords:* Atypicality, Disruption, Citation, Dynamics

2021 MSC: 68U35, 91D30, 68T50

[1] Knowledge Lab, University of Chicago
[2] Department of Sociology, University of Chicago
[3] School of Computing and Information, University of Pittsburgh

* Corresponding authors: Lingfei Wu (liw105@pitt.edu) and James Evans (jevans@uchicago.edu)

**Introduction**

Over the past three centuries, science has exploded in size and recognition as the dominant driver of innovation and economic growth (Jones et al., 2020; Price, 1963). From Derek J. de Solla Price's mid-20th Century scholarship demonstrating extreme inequality in scientific article citations (Price, 1965) and Eugene Garfield's establishment of the Science Citation Index enabling identification of the most "important" papers on a given topic, citation number or "impact" has become the dominant method for quantitatively evaluating researchers and their work. This fixation on citation impact, however, has led to unintended consequences (Hicks et al., 2015). The value of citation impact for careers and institutional allocations has led scientists to make choices that inexpensively optimize the metric without its attendant quality, driving down the value of its index of generalized research quality (Campbell, 1979; Goodhart, 1975; Lucas, 1976). For example, research demonstrates that scientists are rewarded by publishing in fashionable areas of science where citations are abundant (Foster et al., 2015; Rzhetsky et al., 2015), making their work relevant for citation by other contemporary researchers working on those topics. This has contributed to log-jams in science where scientists crowd together along the scientific frontier (Azoulay et al., 2018), driving down the relationship between short-term impact and long-term influence.

Alternative measures have been proposed to reveal the distinct character of scholarly work beyond popularity. These include pioneering work to identify knowledge relevance through bibliographic coupling (Kessler, 1963), co-citation (Small, 1973), or keyword overlapping (Milojević et al., 2011), and initiatives to understand the different function of citations by analyzing their word context, so as to trace the unfolding drama of scientific debate and advance (Biagioli, 2018; Jurgens et al., 2018; Zhang et al., 2013). Two recent, prominent metrics have arisen that highlight work generating new combinations and directions, contributing to the "creative destruction" of science, whereby new scientific ideas and approaches supplant the old (McMahan et al., 2021). Novelty has been assessed in many ways (Foster et al., 2021), but one high-profile approach, *Atypicality*, models how research draws upon unusual combinations of prior research in crafting their own contributions (Uzzi et al., 2013). *Disruption* models how research comes to eclipse citations to the prior work on which it builds, becoming recognized as a new scientific direction (Funk et al., 2016; Wu et al., 2019). In this paper, we unpack the complex, temporally evolving relationship between atypicality, citations, and disruption. We show how atypicality increases at the expense of short-term citations, but anticipates works that will become "sleeping beauties" (He et al., 2018; Ke et al., 2015; van Raan, 2004), accumulating surprising attention and citation impact over the long run to disrupt science. We also reformulate atypicality as distance of co-cited journals in knowledge embedding space inscribed by an embedding model, which offers new theoretical insight into the dynamic frontier of science: how yesterday's novelty forms today's scientific conventions, which condition the surprise of tomorrow's breakthroughs (Fleming et al., 2001; Sorenson et al., 2006).

**Atypicality: Science as a Recombinant Process**

Scientific discoveries and technological inventions do not appear *ex nihilo*. They are built from combinations of existing knowledge and technological components (Brian Arthur, 2009; Schumpeter, 2018; Singh et al., 2010; Uzzi et al., 2013). Recent research has explored the recombination of knowledge entities, including keywords (Hofstra et al., 2020), chemical and biological entities (Foster et al., 2015), and patent classes (Youn et al., 2015) to produce novel scientific advances. For example Hofstra and colleagues identified papers connecting concepts previously viewed as separated or irrelevant in literature

(2020), including how Lilian Bruch connected "HIV" with "monkeys" to introduce HIV's origins in nonhuman primates (Sarngadharan et al., 1984), or how Londa Schiebinger linked "masculinity" to "justify" in pioneering academic studies of gender bias (1991). Milojević and coauthors analyzed combinations of words and phrases from paper titles to elucidate the cognitive content of Library and Information Science (LIS), documenting how its cognitive landscape has been reshaped by the emergence of new information technologies such as the Internet, and the retirement of old ones (Milojević et al., 2011). Much recent work constructs knowledge spaces from combinations of knowledge components, such that the combination of distant elements within that space strongly indicates combinatorial novelty (Gebhart et al., 2020; Shi et al., 2019).

Brian Uzzi and colleagues created a prominent score that captures how a paper deviates from the norm of science by building on "atypical" references, where a pair of journals are determined to be "atypical ($z < 0$)" if they are less likely than random to be co-cited in the existing literature (Uzzi et al., 2013). The atypically of a paper is characterized by the distribution of z-score across all pairs of journals in the references. The 10$^{th}$ percentile of the z-score distribution yields a measure that stably approximates the maximum atypicality, or minimum typicality in combining scientific sources ($z_{min}$), and its 50$^{th}$ percentile yields a measure of average typicality ($z_{median}$). A paper may be highly typical ($z_{median} > z_{min} > 0$), mixing atypical with typical references ($z_{min} < 0 < z_{median}$), or highly atypical ($z_{min} < z_{median} < 0$). Uzzi and his colleagues found that mixing atypical with typical references best predicts the likelihood that a paper will become highly cited (Uzzi et al., 2013). We will show here, however, that the probability for a paper to disrupt rather than consolidate science peaks when it makes the bold move to be highly atypical.

**Disruption: Science Advances in Steps or Leaps**

Scientific work plays distinct roles in the unfolding evolution of science. Research that aims to push the frontier of knowledge differs from that which defends and extends existing theories or solves applied problems (Hicks et al., 2015). As in science, technological change may either consolidate existing knowledge and reinforce established trajectories, or destabilize past achievements and create new paths (Brian Arthur, 2009; Dosi, 1982). This dichotomy reflects a fundamental tension identified by many scholars under different names: "conformity vs. dissent" (Polanyi, 1962), "succession vs. subversion" (Bourdieu, 1975), paradigm "deepening vs. changing" (Ahuja et al., 2014; Dosi, 1982), "enhancing vs. destroying" (Tushman et al., 1986), "exploitation vs. exploration" (March, 1991), "relevance vs. originality" (Whitley, 2000), "conventionality vs. novelty" (Uzzi et al., 2013), "tradition vs. innovation" (Foster et al., 2015), "destabilization vs. consolidation" (Chen et al., 2021), or path "deepening vs. breaking" (Garud et al., 2010; Karim et al., 2017). Some scholars have argued that the two types of science and technology are fundamentally incompatible and present an essential tension in which scientists must trade one for the other (Bourdieu, 1975; March, 1991), producing different career outcomes for those that undertake them and contributing distinct possibilities for scientists that follow (Foster et al., 2015). Others have argued that these two forces take turns characterizing the history of science, switching between "normal vs. revolutionary" periods (Kuhn, 1962). With all of the scholars listed above, we argue that these strategies are complementary; both are critical for sustained advancement in science.

To this end, recent work has sought to explore how science develops vs. disrupts prior science over time (Funk et al., 2016; Wu et al., 2019). The intuition behind the proposed disruptive D-measure is straightforward: if subsequent papers that cite a paper also cite that paper's references, then the focal paper can be seen as consolidating the prior knowledge upon which it builds. If the converse is true—future papers citing a focal paper ignore its acknowledged forebears—they are recognizing that paper as disruptive, creating an unanticipated new direction for science. The D-score of a focal paper is calculated as the difference between the fraction of these types of subsequent, citing papers. A paper may largely eclipse prior work by introducing distinct ideas ($0 < D < 1$), develop and promote existing theories by providing supportive evidence ($-1 < D < 0$), or balance both ($D = 0$). Through the lens of D-score, the BTW-model paper (Bak et al., n.d.) that discovered the "self-organized criticality" property of dynamical systems, one of the most prominent patterns in complexity science, is among the most disruptive papers ($D = 0.86$, top 1%). In contrast, the article by Davis et al. (Davis et al., 1996) that first observed Bose-Einstein condensation in the lab, a property hypothesized nearly three quarters of a century before (Bose, 1924), is among the most developing ($D = -0.58$, bottom 3%). D-scores highlight the distinct nature of knowledge created by these two papers that cannot be captured with citation counts—both papers received over 8,000 citations, according to Google Scholar. The D-scores of 20 million Microsoft Academic Graph papers (1830-2021) are provided for public use at https://lingfeiwu.github.io/smallTeams/. Since publication of the disruption index, many papers have recommended adjustments (Bornmann et al., 2020, 2019; Chen et al., 2021) or extensions (Leahey et al., 2021; Leydesdorff et al., 2021).

**The Delayed Recognition of Scientific Novelty**

The delayed recognition of papers has been discovered for decades, but it was not until recently that the importance of this phenomenon in science was recognized. Glänzel and Garfield observed that most papers receive most of their citations within the first three to five years of publication (Glanzel et al., 2004), except for a negligible fraction of outliers—0.01% according to their study—which experience a burst of attention after ten years. Recent studies of large-scale citation graphs have confirmed the scarcity of delayed recognition papers (Wang et al., 2013; Yin et al., 2017), but these outliers, named "sleeping beauties," may not be so rare (He et al., 2018; van Raan, 2004). (Ke et al., 2015) proposed a "sleeping beauty index" (SBI), a non-parametric measure, and reported that papers with top 0.1% SBI demonstrated a clear pattern of delayed recognition, ten times larger than what Glänzel and Garfield suggested. One possibility is that Glänzel and Garfield only analyzed papers published before 1980, and thus missed the opportunity to discover a majority of "sleeping beauties" papers "awakened" after that. But there is an elephant in the room that Glänzel and Garfield ignored. Papers garnering belated recognition may be rare, but this does not mean they are unimportant. On the contrary, it is likely that these papers are too novel to be recognized immediately and their importance must unfold over time. Following this rationale, we ask the following questions concerning the social mechanism through which novel papers are recognized:

**Question 1**. How often does a novel paper successfully create a new direction and disrupt science? Stated in a different way, do novel inputs predict disruptive outcomes? We anticipate that paper novelty should be positively correlated with future disruption rather than development, as novel combinations are more likely to depart from existing trajectories and open new paths (Fleming et al., 2007; Lee et al., 2015; Tushman et al., 1986). But does disruption necessarily grow from unusual combinations? Can one create a new direction by citing interlinked sources and fighting against consensus? Rarely. The physicist Max Planck made the sharp observation restated as "science progresses one funeral at a time", an idea widely cited by Thomas Kuhn, Paul Feyerabend and other science and technology studies (STS) scholars. Old perspectives die not from new arguments, methods and evidence, but from the marginalization (H. M. Collins, 2000) and death (Azoulay et al., 2019) of those who hold them. This suggests that ignoring old arguments may be more likely to generate new directions in science than disputing them. As such, we posit that disruption grows much more from novel than conventional combinations of prior ideas and literatures.

**Question 2.** If novel papers indeed disrupt science, how long does this process take? Based on our earlier discussion, we anticipate that novel papers are more likely to be "sleeping beauties" and accumulate citations over the long run, as new findings that contradict traditional wisdom (Cole, 1970) or appear "before their time" (Garfield, 1980; van Raan, 2015) may face resistance from contemporary ideas and their defenders, achieving delayed recognition from other fields only as knowledge spills over (Ke et al., 2015). If novel papers are indeed more disruptive as theorized above, they are likely to become so by attracting citations over the long term. This is because the D-score of a focal paper is calculated as the difference between two types of papers: papers that only cite the focal paper but not its references and those citing both. These two types may not appear with the same likelihood over time: short-term citations will likely cite both, while long-term citations, further in time from the work a focal work cited, will disproportionately cite the focal paper alone. This would explain the finding that D-score is associated with the "sleeping beauty" index (Wu et al., 2019). In this paper, we will unfold the temporal dynamics of disruption and answer this question.

**Question 3.** How does the landscape of scientific novelty evolve? We predict that landscapes of novel opportunity in science evolve continuously, with every new finding and claimed association. While analysis of recombinant ideas has been used to quantify the novelty of individual papers in historical context, similar approaches have not yet explored the changing context itself. Here we seek to understand how yesterday's novelty forms today's scientific conventions, which condition tomorrow's breakthroughs. For example, Mark Granovetter's classic paper on social networks published in 1973, "The Strength of Weak Ties", cited both physical science journals (e.g., *Bulletin of Mathematical Biophysics*) and sociological journals (e.g., *American Sociological Review*). This combination was much more novel in the 1970s than today, partly as a result of the success of this early work (Castellano et al., 2009). Recent advances in semantic analysis such as the word2vec model (Mikolov et al., 2013) provide a powerful tool to reformulate paper atypicality defined by (Uzzi et al., 2013) as distance across the underlying, continuous space of knowledge in which conventional and novel combinations are continuously redefined. Within these spaces, "structural holes" (Burt, 2004) or sparse regions that separate distinct communities and fields, emerge and collapse like whirlpools in the ocean. We propose to model and visualize knowledge space over time to reveal changes in novelty and facilitate understanding regarding how the scientific frontier evolves.

**Data and Method**

**Data.** We investigate impact, novelty and disruption using the Microsoft Academic Graph (MAG), which includes 87,860,684 journal articles published 1800-2020 and 1,042,590,902 citations created by these articles. We calculated two variables for each of the 35,431,832 journal articles that have both citations and references, including D-score for disruption (higher is more disruptive) and $z_{median}$ for typicality (higher is more conventional). Our analysis of typicality, citations, and disruption covers multiple cohorts,

including cohorts from 1970 (87,475 papers), 1980 (176,826 papers), 1990 (318,914 papers), and 2000 (591,653 papers). These papers have 21 references on average. The average number of citations to these papers is 32. The construction of the journal embedding space is based on the co-citation of 2,429 journals in 1970 and 8,009 journals in 2000.

**Methods.**

**Calculating the z-score of novelty.** Uzzi et al. (2013) defined the z-score for a pair of journals co-cited in an article as more or less typical with $z_{ij}$,

$$z_{ij} = (obs_{ij} - exp_{ij})/\sigma_{ij} \quad (1)$$

where $i$ and $j$ are journals, $obs_{ij}$ is the empirical frequency that these two journals co-cited across research articles (Uzzi et al., 2013) and $exp_{ij}$ is the expected frequency of the co-citation. $exp_{ij}$ is calculated from citation random shuffling, in which two citations are randomly selected to exchange the papers to which they are attached, given that the papers are published in the same year. In this way, two variables remain unchanged, including the length of references for each paper and the temporal distribution characterizing their references.

**Calculating the D-score of disruption.** The Disruption, $D$, of a focal paper, can be calculated as the difference between the fractions of two types of subsequent papers,

$$D = p_i - p_j = \frac{n_i - n_j}{n_i + n_j + n_k} \quad (2)$$

where $p_i$ is the fraction of papers that only cites the focal paper but not its reference and $p_j$ is the fraction of papers that cites both. A paper may disrupt earlier research by introducing new ideas that come to be recognized independent from the prior work on which it builds ($0 < D < 1$), develop existing research by providing supportive evidence or extensions that come to be recognized as developments of prior work ($-1 < D < 0$), or remain neutral, keeping in balance the disruptive and developmental character of its contribution ($D = 0$). D-score may change over time due to the temporal evolution of the two types of subsequent, citing papers. To calculate a stabilized disruption score, we used the longest time window available in the MAG dataset from the year of publication for each paper to 2018. In section 1 of our findings, we explore the temporal dynamics of $D_t$, i.e., how D-score changes with time, for two field-definitive studies in biology, the paper on DNA by Watson and Crick (Watson et al., 1953) and the paper on RNA by Baltimore (Baltimore, 1970) and also four cohorts of papers published in 1970, 1980, 1990, and 2000, respectively.

**Reformulating z-score atypicality as distance in knowledge space.** The z-score of atypicality is deeply related to a common measure in information science, the Pointwise mutual information (PMI) between two items. Indeed, we can rewrite PMI into a z-score-like form

$$PMI_{ij} = log_2(\frac{P_{ij}}{P_i \times P_j}) = log_2(P_{ij}) - log_2(P_i \times P_j) = log_2(obs_{ij}) - log_2(exp_{ij}) \quad (3)$$

where $P_i$ and $P_j$ are the probabilities that $i$ and $j$ appear independently, respectively, and $P_{ij}$ is the joint probability. The hidden connection between PMI and z-score permits defining and measuring atypicality as the distance on latent semantic spaces obtained through an embedding model, such as the popular skip-gram word2vec model, which has been demonstrated to preserve semantic compositionality within word vectors sufficient to perform at human level on semantic analogy problems (*a* is to *b* as *c* is to \_\_\_?) (Mikolov et al., 2013). Word embedding models have inspired a wide range of item-context embedding models beyond words, ranging from images (Xian et al., 2016) and audio clips (Xie et al., 2019) to graphs (Grover et al., 2016; Perozzi et al., 2014) and academic journals (Miao et al., 2021; Peng et al., 2020; Tshitoyan et al., 2019).

In an embedding model, each item is represented as a vector in shared vector space. For example, in a word embedding, words sharing similar contexts within the text will be positioned nearby in the space, whereas words that appear only in distinct and disconnected contexts will be positioned farther apart. The same holds for journals embedded as a function of their co-citation within reference lists. Consider the structure of the descriptive problem that embeddings attempt to solve: how to represent all items from a dataset within the *k*-dimensional space that best preserves distances between *n* items (e.g., journals) across *m* contexts (e.g., article reference lists). The solution, is a *n*-by-*k* matrix of values, where $k \ll m$. Early embedding approaches used singular-value decomposition (SVD) to factorize this item-context matrix, where contexts were large and nondiscriminating (e.g., entire documents of thousands or tens of thousands of words), but SVD placed strict upper limits on the number of contexts they could factorize. Neural embeddings use heuristic optimization of a neural network with at least one "hidden-layer" of *k* internal, dependent variables, enabling factorization of much larger item-context matrices constructed from vast numbers of arbitrarily local item contexts (very large *m*). In this way, PMI is formally equivalent to the inner product of two vectors representing items within a latent semantic space (Levy et al., 2014). Specifically,

$$emb_{in-i} \bullet emb_{out-j} = PMI_{ij} - log_2 Neg \qquad (4)$$

where $emb_{in-i}$ is the item embedding of *i* and $emb_{out-j}$ is the context embedding of *j*. *Neg* is the number of negative samples per positive (actual item-context) sample. In sum, the inner product between journal vectors in an embedding space is a computationally efficient proxy for the z-score. In section 3 of our findings, we train journal vectors across time periods to visualize and compare the changing landscape of novelty in science.

**Findings**

**1. Novel papers are more likely to disrupt existing literature**

We find that typicality ($z_{median}$) and disruption (D-score) are negatively associated (Pearson correlation coefficient -0.05, p-value < 0.001). Papers integrating unusual combinations of literature come to be seen as disruptive by a disproportionate number of subsequent papers that only cite those novel papers but not their references. In comparison, papers drawing upon typical combinations of references are deemed as developing prior approaches by the majority of following papers that cite those papers in context with their references—as extensions.

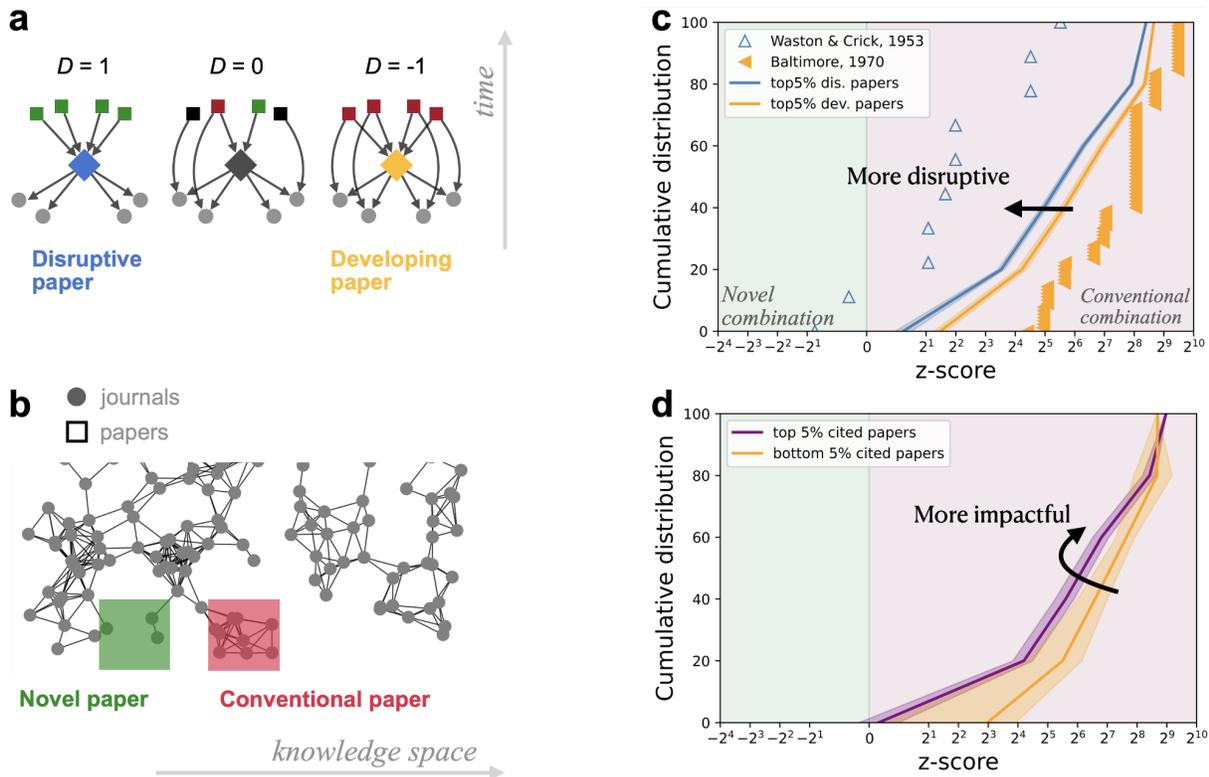

**Figure 1. Novel papers disrupt, conventional papers develop.** (a) Simplified illustration of disruption. Three citation networks comprising focal papers (colored diamonds), references (grey circles) and subsequent work (grey squares). Subsequent work may cite the focal work (green squares), both the focal work and its references (red squares) or just its references (black squares). Disruption, $D$, of the focal paper is defined by the difference between the proportion of type $i$ and $j$ papers $p_i - p_j$, which equals the difference between the observed number of these papers $n_i - n_j$ divided by the number of all subsequent works $n_i + n_j + n_k$. A paper may be disrupting ($D = 1$), neutral ($D = 0$) or developing ($D = -1$). Figure recreated from (Wu et al., 2019). (b) Simplified illustration of novelty. A paper may cite journals of weak (green, $z < 0$) or strong ties (red, $z > 0$). (c) Cumulative distributions of z-scores for two exemplary papers: the DNA paper by Waston and Crick ($D = 0.96$, top 1% disruptive) and the RNA paper by Baltimore ($D = -0.47$, top 1% developing). For all 87,475 papers published in 1970, we selected the most disruptive (top 5%) and developing (top 5%) papers, then calculated their average cumulative distribution of conventionality (displayed in blue and orange, respectively). Z-median for disruptive papers is significantly different from those of developing papers (the Kolmogorov–Smirnov statistic $D = 0.14$, $p < 0.001$). The same conclusion holds for z-min ($D = 0.09$, $p < 0.001$). (d) The cumulative distributions of the z-score of high-impact (purple, top 5% citations) and low-impact (yellow, bottom 5% citations) papers selected from all of the 87,475 papers published in 1970. Z-median for high-impact papers is significantly different from low-impact papers (the Kolmogorov–Smirnov statistic $D = 0.15$, $p < 0.001$). Note that Panels c and d are plotted using the "symlog" (which means symmetrical log) function from the "matplotlib" library of Python.

Fig.1 presents the association between typicality and disruption with two representative cases after illustrating the calculation of D- and z-scores. In Fig. 1c, each paper is characterized by the distribution of the z-score for all pairs of journals in the references. A high, positive z-score (distribution shifting to the right end on the x-axis) is a signature of typical combinations of journals on established topics within a field, whereas a low, negative z-score (distribution shifting to the left end on the x-axis) implies unusual combinations of journals that span fields to create new topics. More specifically, The 10$^{th}$ percentile of the z-score distribution yields a measure that stably approximates the maximum atypicality, or minimum typicality in combining scientific sources ($z_{min}$), and its 50$^{th}$ percentile yields a measure of average typicality ($z_{median}$). A paper may be highly typical ($z_{median} > z_{min} > 0$), mixing atypical with typical references ($z_{min} < 0 < z_{median}$), or highly atypical ($z_{min} < z_{median} < 0$). We find that only a small fraction of papers (2.3%) are highly atypical and there are less atypical papers over time, with a nearly threefold decrease from 3.9% in 1970 to 1.4% in 2000.

Uzzi and colleagues discovered that mixing conventional with atypical references best predicts the likelihood that a paper will become highly cited (Uzzi et al., 2013). This finding is confirmed in Fig. 1d. However, Fig. 1c shows that the chance of disrupting rather than consolidating science peaks when a paper takes the bold move to be highly novel. A highly novel paper is much more likely—nearly two times (61% vs 36%), to disrupt science than conventional papers. The average z-score distribution of the most disruptive (top 5% D-score) versus developing (bottom 5% D-score) papers significantly deviate from one another as evidenced by Kolmogorov-Smirnov tests (see the caption of Fig. 1). The former shifts to the left and the latter shifts to the right, indicating the alignment between novelty and disruption; conventionality and development. The correlation between atypicality and disruption holds for a majority of fields, but the effect is more significant in "artificial" than "natural" sciences (Simon, 2019). In computer science and engineering (the higher average D-score as presented in Fig. 3 in (Wu et al., 2019), atypicality is more likely rewarded in disruption, evidenced by the larger difference between novel versus conventional groups in the fraction of disruptive citations (Table S1). By contrast, for stable sciences such as biology, chemistry, and physics, which remain difficult to disrupt (low average D-score), atypicality is more weakly related to disruption.

The fundamental difference between dynamics revealed in Fig. 1c and 1d should not be underestimated. Scientific advance is constrained by an essential tension between "tradition vs. innovation" (Foster et al., 2015): in most cases, new ideas must be introduced in connection with relevant, old ideas to enter the canon of scientific knowledge (Collins 2009; Chu and Evans 2021). This permits two types of strategies for individual scientists to effectively contribute. One can prioritize tradition by selecting an established theory and adding new, supportive evidence. This strategy, characterized by the "clockwise rotation" and decreased slope of the z-score distribution (Fig. 1d), an operation suggested by the black, curved arrow, maximizes the chance a paper will achieve "hit" status (Uzzi et al., 2013). Alternatively, one can prioritize innovation by selecting an underdeveloped topic lacking consensus. In the space of z-scores, this strategy corresponds to "left shifting" the cumulative distribution (Fig. 1c), which results in a higher likelihood of being disruptive.

A question that remains is when atypical papers disrupt science, are they cited as sources of the new concepts they contribute? To answer this question, we selected 887 scientific keywords identified by MAG to create 887 groups of papers that contain them. We then separate the most cited paper from the other papers in each group, and compare these two types of papers in atypicality and disruption. We anticipate that by selecting the most cited paper, we can reveal the "center" of the scientific consensus. For example, among all 22 papers containing the keyword "the market for lemons," the paper "The Market for 'Lemons': Quality Uncertainty and the Market Mechanism (Akerlof, 1970)," was the most highly cited. This paper was indeed among the first discussing the consequence of information asymmetry on markets between buyers and sellers (D = 0.99, $z_{median}$ = 14). We find the "center" papers are more disruptive and atypical than other papers, supporting the assumption that atypical and disruptive papers were recognized as the source of the new concept they contributed.

To obtain a more intuitive understanding of the complex dynamics characterizing "creative destruction" in science, whereby new scientific ideas and approaches supplant the old (McMahan et al., 2021), we highlight two papers, one on the double helix structure of DNA by Waton and Crick (Watson et al., 1953) called the "DNA" paper hereafter, and another on "RNA-dependent DNA Polymerase" by David Baltimore called the "RNA" paper hereafter. The two papers are similar in many ways: both are highly-cited, field-definitive work by distinguished biologists later awarded the Nobel Prize in Physiology or Medicine. However, the z-score distributions reveal their distinct approaches to integrate prior knowledge (Fig. 1c). Baltimore's article reviewed papers published in conventional biology venues such as *Virology* and *Biochemical and Biophysical Research Communications* ($z = 51$) and hypothesized that genetic information could transfer bidirectionally between DNA and RNA. At the time of this paper, information transfer from DNA to RNA was well studied, and Baltimore was not the only scholar proposing to test the reverse influence from RNA to DNA. Actually, Baltimore's paper was published back-to-back with Howard Temin's paper (Mizutani et al., 1970) on the same topic in the same issue of *Nature*. The bidirectional influence between DNA and RNA represents the "adjacent possible" described by Kauffman (Kauffman, 1996), which articulates new ideas or discoveries that extend from prior science, a single step from present understanding as "low hanging fruit", easily reached, which, unsurprisingly, triggers intense competition. Back-to-back publications by Baltimore and Temin were like the race to the South Pole between Britain's explorer Robert Scott and Norway's Roald Amundsen. Unlike the explorers' race, which ended in victory for Amundsen and tragedy for Scott, the discovery of DNA-RNA's mutual influence became a shared and widely celebrated success treated as a confirmation of the underlying claim. In 1975, only five years after the paper's publication, Baltimore and Temin shared the Nobel Prize in Physiology or Medicine. This timely appreciation itself speaks for the adjacent, developing nature of that discovery, evidenced by its low D-score (D = -0.47, bottom 1% disruption, or top 1% development).

In comparison, Watson and Crick's paper cited prior literature published in diverse journals across fields, including *Journal of Geophysical Research* and *Journal of Chemical Physics* ($z = -26.7$), *Canadian Journal of Chemistry* and *Quarterly Journal of the Royal Meteorological Society* ($z = -5.7$), proposing that double-stranded DNA of helical structure is the genetic material. This paper was ahead of time. When published, there was not yet a consensus on the identity of genetic material—proteins seemed a better bet. Moreover, few could foresee its influence into the future; how the double helix shed light on almost every aspect of modern biology and medicine for decades to come, ranging from the migration of human populations and cancer-causing mutations in tumors to the diagnosis and treatment of rare congenital diseases. Watson and Crick received delayed recognition of the Nobel Prize in Physiology or Medicine in 1962, ten years after the paper was published—an enduring wait twice longer than Baltimore's work despite its greater impact in transforming the future of biology and offering the non-academic world an icon of scientific work—the double helix (Ferry, 2019). This delayed acknowledgement footnotes the pioneering, disruptive nature of that discovery, evidenced by its high D-score (D=0.96, top 1% disruptive).

## 2. Novel papers are more likely to become "sleeping beauties" and accumulate citation impact over the long run

Going back to the moment of publication, can one have foreseen the accelerated acknowledgment of Baltimore's contribution and retarded recognition to Watson and Crick? Could we predict them from the typicality of the former ($z_{median} = 266.3$) and atypicality of the latter ($z_{median} = 4.8$), which can be derived at the point of publication? In these cases, and millions of others published over the following decades, we document that novelty results in delayed impact. Creative explorations that travel more than a step beyond the adjacent possible (Monechi et al., 2017) are inaccessible to the majority of scientists upon publication, but come to make up the pool of possibilities that are verified, appreciated, or reformulated and used to advance science over the long run.

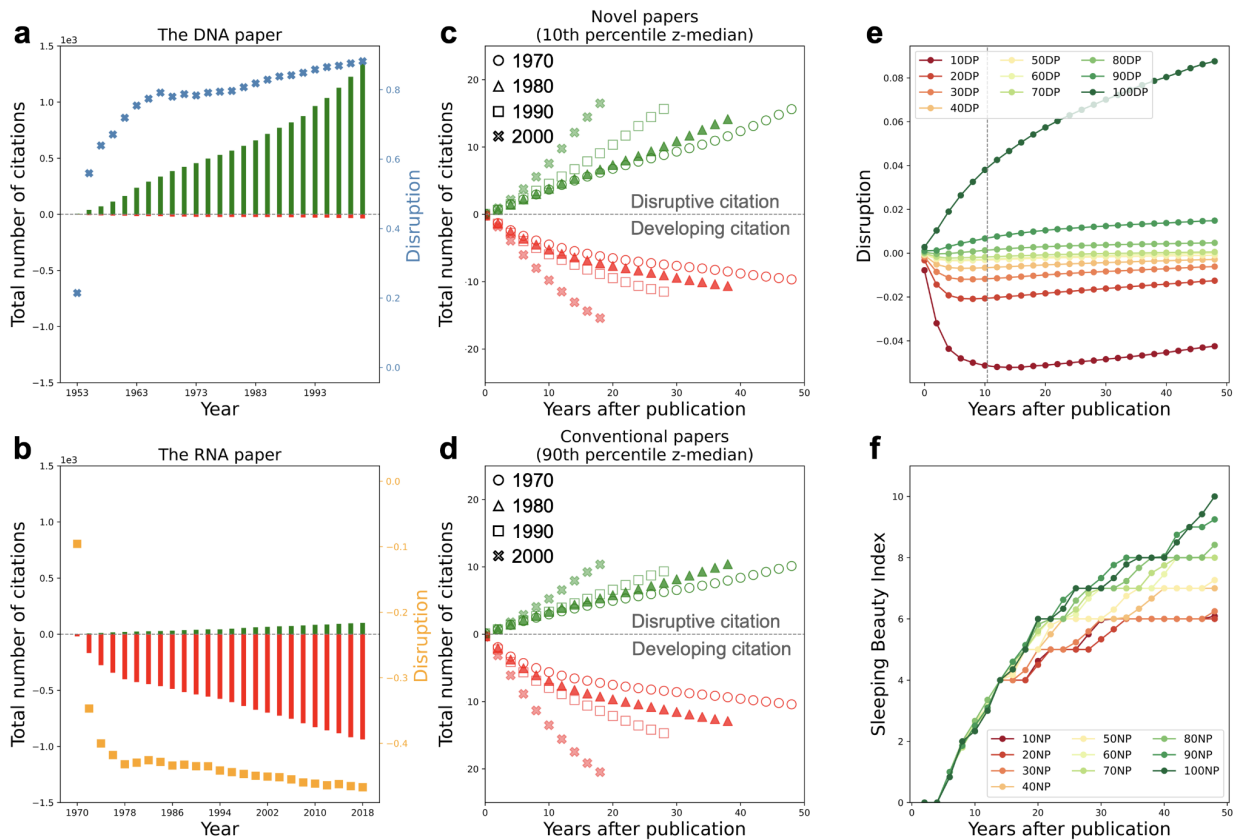

**Figure 2. The disruption of novel papers increases over time.** In Panels a-b, we present the temporal evolution of disruption scores for the DNA paper by Waston and Crick (a) and RNA paper by Baltimore (b), and the total number of disruptive (green) and developing (red) citations to the DNA paper and RNA paper over time, respectively. This analysis is extended to four generations of papers, including the cohort of 1970 (87,475 papers), 1980 (176,826 papers), 1990 (318,914 papers), and 2000 (591,653 papers). For each cohort, we select the most novel (top 10% z-median, Panel c) and conventional papers (bottom 10% z-median, Panel d) and plot the average total number of citations over time. Statistical tests on the asymmetry between the two types of citations were reported as follows. The t-test of the difference between novel and conventional papers in the disruptive-citation fraction from the last year of analysis (2018) is significant for all cohorts, including 1970 (t-statistic = 23.85, p < 0.001), 1980 (t-statistic = 33.04, p < 0.001), 1990 (t-statistic = 52.78, p < 0.001), 2000 (t-statistic = 88.76, p < 0.001).
In Panel e, from all the 87,475 papers published in 1970, we break them into ten groups by disruption percentiles (DP) and plot the average disruption score of papers within the group against years. The curves are colored by DP. The vertical grey line shows that it typically takes 10 years for the D-score to stabilize (the earliest time for a paper to reach 80% of its final D-score). In Panel f, from all 87,475 papers published in 1970, we break them into ten groups by novelty percentiles (NP) using z-median, and plot the median Sleeping Beauty Index (SBI) of papers within the group against years. In the original version, each paper has only one SBI calculated over the lifecycle of citations (Ke et al., 2015). Here we calculate and track the temporal evolution of SBI for each two-year time window. Average curves for low-novelty papers flatten within a decade, showing no delayed burst of citations to those papers. Average curves for high-novelty papers continuously increase after a decade, showing that new, larger bursts still happen after a long wait, i.e., long-term citations pay off the lengthy wait.

We examined the temporal evolution of disruption and citations for these two, archetypal papers (Fig. 2a-b). We find that the D-score of the DNA paper increased nearly monotonically from 0.2 in 1953 to nearly 0.8 ten years after publication, steadily increasing to 0.96 in 2018, whereas the D-score of the Baltimore paper has been negative since publication in 1970, and decayed rapidly to -0.45 within five years of publication, barely changing in the subsequent four decades (D = -0.47 in 2018, bottom 1%). This seems to be a general pattern we confirmed from many other cases: the D-score of developing papers converges quickly within five years, but that of disruptive papers increases after a decade or longer (Fig. 2e).

For both the DNA and RNA papers, we unpacked two kinds of citations: disruptive citations from subsequent papers that only cite the focal paper but not its references (green curves in Fig. 2a), and developing citations from subsequent papers that cite both (the red curves in Fig. 2b). We find a "taking off" pattern in the DNA paper—disruptive citations increase steadily following paper publication, deviating from developing citations, which decline exponentially after a short peak and follow the widespread pattern of citation decay (Wang et al., 2013). Disruptive citations contribute to long-term impact more than developing citations. In comparison, citation impact of the RNA paper is increasingly dominated by developing citations, reflecting the stabilizing consensus on its developing contribution within biology.

Long-term impact for atypical papers is confirmed as a general pattern when we scale the data (Fig. 2c-d). We select the most novel (top 10%) and conventional (bottom 10%) papers by z-median including the cohort of papers published in 1970 (87,475 papers), 1980 (176,826 papers), 1990 (318,914 papers), and 2000 (591,653 papers) then compared the difference between disruptive (green data points in Fig. 2c-d) and developing (red data points in Fig. 2c-d) citations. Atypical papers that integrate surprising combinations of literature to create new ideas accumulate long-term impact by attracting both disruptive and developing citations, with the relative fraction of the former over the latter amplifying over time (Fig. 2c). This pattern reverses in the citation dynamics of conventional papers, wherein the relative fraction of developing citations increases faster than that of disruptive citations (Fig. 2d).

To verify the long-term impact of atypical papers, we calculate the Sleeping Beauty Index (SBI) (Ke et al., 2015). A paper with a high SBI will receive few citations upon publication, followed by a later burst tracing a convex curve. By contrast, a paper with a low SBI will receive many citations following publication and fewer later tracing a concave cumulative distribution. Atypicality and SBI are positively correlated (Pearson correlation coefficient equals 0.08 on the log-log scale, p-value < 0.001). We also calculated SBI over time to examine the chance that larger bursts occur after a long wait, i.e., whether the long-term citations compensate for lengthy waiting times and drive up SBI over time. We find that curves of conventional papers flatten within a decade, implying no delayed burst of citation attention. In contrast, the citation curves of atypical papers continuously increase after a decade, revealing long-term citation pay-off and scientific influence after lengthy waiting times (Fig. 2f).

**3. Reformulating atypicality as distance in knowledge space to map the moving frontier of science**

We first demonstrate that atypicality can be reformulated for computational efficiency and dynamism as the distance between journals in embedding spaces built from co-cited journals (see Methods for details). Recent advances in natural language processing (NLP) for semantic analysis, such as the word2vec manifold learning model (Mikolov et al., 2013), provides us the computational tools needed to reconstruct knowledge spaces. With them, we can extend the z-score as a measure of continuous distance across embedding space. Levy and Goldberg analytically proved that PMI, a revised z-score, equals the distance between vectorized items embedded in latent spaces as calculated by their inner product (Levy et al., 2014). In this way, knowledge embedding spaces learn scientific conventions, which can be used to assess and direct exploration of the scientific frontier (Tshitoyan et al., 2019). The temporal evolution of these knowledge embedding spaces reveal how yesterday's novelty forms today's scientific conventions, disrupted by tomorrow's breakthroughs.

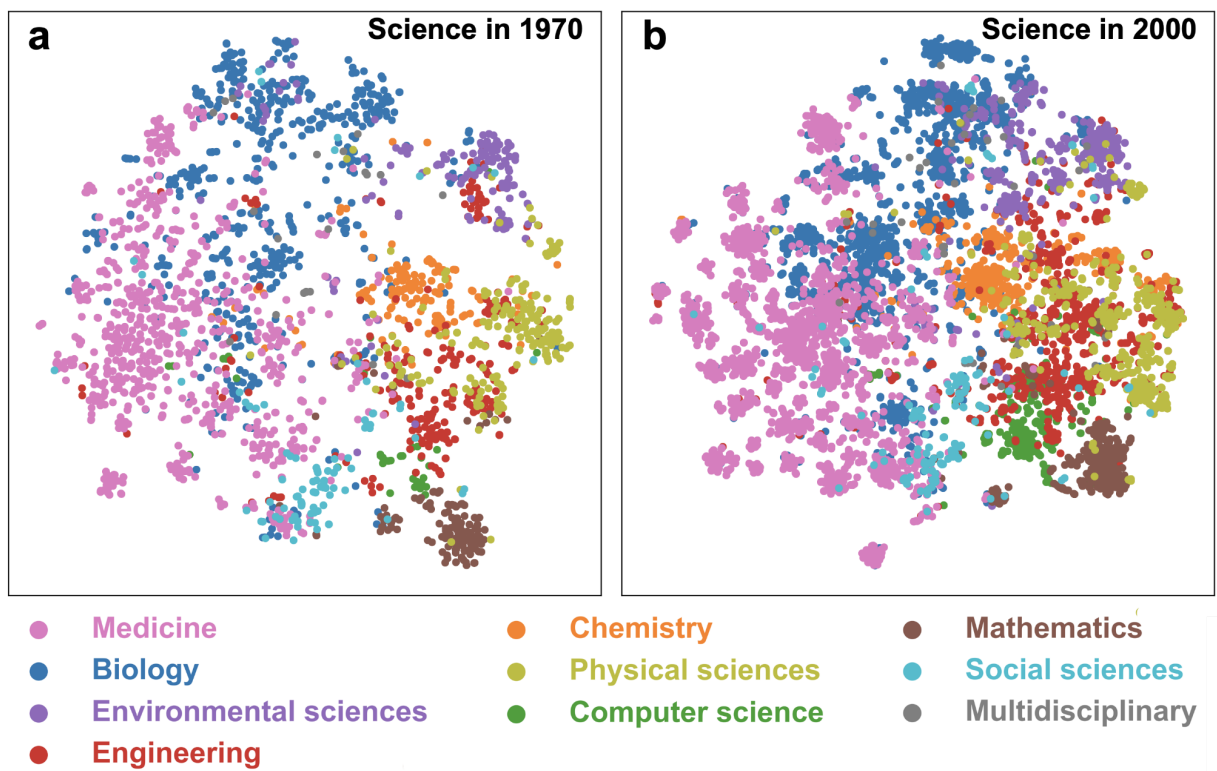

**Figure 3. Knowledge spaces in 1970 and 2000 obtained through journal embeddings.** We constructed two journal embeddings using the 1970 (2,429 journals) and 2000 (8,009 journals) cohort of papers (see Methods for details of the embeddings). Each dot is a journal colored by field. We trained the embeddings using the word2vec Skip-gram algorithm. We used the Gensim package in Python with parameters as follows: embedding dimension = 50, negative sampling size = 5 and window size = 10. We then project the 50-D journal vectors into a 2-D space using the t-SNE algorithm.

To test the association between z-score and embedding space distance, we constructed two embedding spaces from journal co-citation networks in 1970 (2,429 journals) and 2000 (8,009 journals), respectively. We also visualize these embedding spaces by projecting it onto a 2-dimension space with the t-SNE algorithm. We find that the distance between journals in the embedding space reflects their content relevance, as journals from the same field tend to cluster together (Fig. 3 and Fig. S1). This observation is confirmed as we zoom-in with a focus on the regions covered by Mathematics and Computer Science journals. While close to each other, the average distance between journals across the two fields is larger than within a field. The z-score for journal pairs correlates strongly with the inner product between journal embeddings (Pearson correlation coefficient = 0.74, p-value < 0.001), confirming the validity of our z-score reformutation (Fig. S2).

The landscape of novelty is dramatically changing, as revealed in comparison between knowledge spaces from 1970 and 2000 (Fig. 3). One of the most strikingly visible trends is the emergence of dense areas of journals within each field, suggesting the formation of subfields supported by consensus on relevant topics. Clustering occurs within fields, but fields also mix with one another, showcasing the increasing importance of interdisciplinary scientific collaboration (Leahey et al., 2017; Leydesdorff et al., 2021). The change in relative distance between fields also reveals rich trends in shifting atypicality. In the 1970s, a study drawing together social and computer science was highly atypical due to the distance between these two fields, but is far less so in 2000, when these two field are closer to each other after waves of movements that link them, including "social informatics" of the 1980s (Kling, 1999) and "computational social science" in the past decade (Lazer et al., 2009, 2020).

**Discussion**

Citation data should be analyzed in a way that distinguishes different scientific contributions, unlike citation counts, which project all papers onto a single dimension of popularity. Unfortunately, over the past several decades, citation impact and its derivatives (e.g., impact factor and H-index) have come to determine a scholar and institution's viability. This has unintentionally led to a fixation on short-term scientific advances that crowd together along the scientific frontier to predictably yield citations, but which does not propel science forward by generating novel combinations and new alternatives. With the availability of rich data on networks of scientists, institutions, and ideas in concert with advances in natural language processing and increased computational power, we can now represent the high-dimensional character of science with fidelity and expand evaluation metrics about research, researchers, and institutions. This paper has unpacked the complex, temporally evolving relationship between citation impact alongside metrics of atypicality and disruption, which focus on path-breaking contributions that move the scientific frontier.

A paper may choose between two types of research contribution, reflected in the intercorrelation of their references. If references are highly clustered, consisting of frequently co-cited sources (Small, 1973) or pairs of high "typicality"(Uzzi et al., 2013), this implies the existence of strong consensus on the topic (Shwed et al., 2010). The paper contributing to such a topic is more likely to be viewed as a part of an ongoing conversation, and future papers will likely judge it as developing an established direction, citing it together with its references (D < 0, Fig. 4a). By contrast, if the referenced literature is unstructured, only loosely linked or even disconnected (z < 0), this suggests a lack of consensus and underdevelopment of the topic. In this case, the focal paper may creatively and ambitiously integrate scattered ideas and, if successful, is more likely to be recognized as creating a new direction for which future papers will cite the paper alone, without its references (D > 0, Fig. 4b).

## How a new direction for science is created?

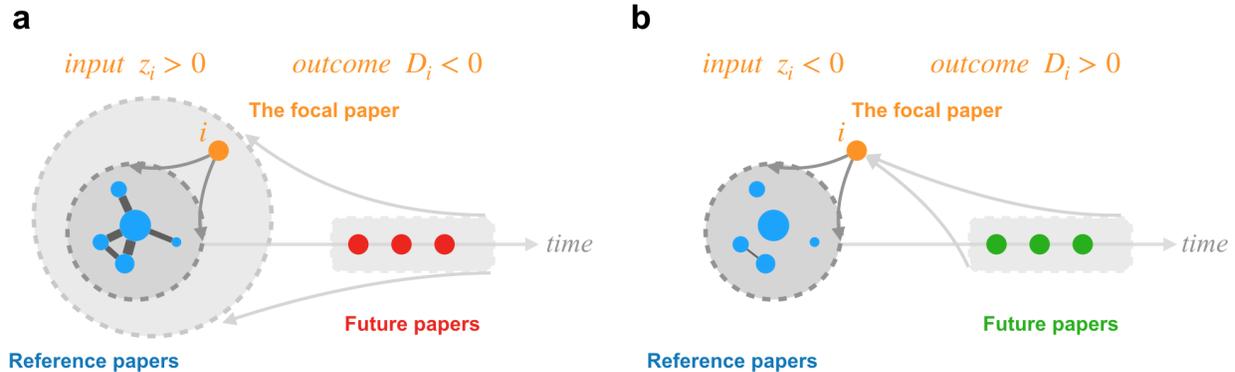

**Figure 4. An illustration of how novelty and disruption are related.** In Panel a, the focal paper draws upon and contributes to literature on a well studied topic, characterized by "clustered" references that have high, pairwise "typicality" ($z > 0$) (Uzzi et al., 2013) as they are frequently co-cited (Small, 1973) and form "strong-ties" (Granovetter, 1973). Network modularity (Newman, 2006) emerges from these strong-ties, reflecting well-established consensus (Shwed et al., 2010). In Panel b, the focal paper identifies an unsolved problem and addresses it by integrating distant literature that is loosely linked or even disconnected ($z < 0$), which implies a lack of consensus. These two kinds of intellectual activities are not only different in input, but generate very different outcomes. The focal paper contributing to a developed topic is more likely to be viewed in future as a part of the ongoing conversation. For this reason, future papers (red circles in Panel a) will judge this paper as developing the broader topic or field (the larger gray circle in Panel a) and cite it together with its references (D < 0). By contrast, the focal paper identifying unsolved problems tends to be viewed as creating a new direction for which future papers (green circles in Panel b) will cite it as a starting point, ignoring its references (D > 0).

We have demonstrated that novel versus conventional science yields contributions that disrupt versus develop past science, respectively, and can be clearly distinguished from citation data. We show how the new D-score and z-score measures are not only useful as novelty metrics for individual papers, but also provide powerful tools to understand how science advances, driven by the essential tension of "tradition vs innovation" (Foster et al., 2015). As illustrated in Fig. S3, if science continues to expand, continuously searching out new topics, a majority of papers will be disruptive (D = 1), but no consensus will be achieved and no tradition will form. Alternatively, if all new papers cite existing clusters of papers, a majority will develop (D = -1), consensus will be established, but no new ideas will be possible as the knowledge space collapses to a closed, crystalline system. Sustainable advance requires that science balance tradition and innovation.

We found that nearly 67% papers develop prior science (D < 0), revealing the conservative nature of most scientific activity. The slow path to acceptance for novel and disruptive research contributions points to sustained resistance against radically new ideas. This underscores the history of how many significant breakthroughs of modern science were initially rejected or ignored, sometimes for decades. Consider Darwin's theory of evolution introduced in 1859; the atom conception proposed by Ludwig Boltzmann in the 1870s, the Continental Drift model formulated in 1912 by Richard Wegener; the Big Bang theory of the the origin of the universe formalized by Georges Lemaître in the 1920s, and the gravitational wave theory by Albert Einstein in 1916. Resistance to more recent science includes denial of documented hazards from tobacco and DDT, ozone depletion, and climate change (Oreskes et al., 2011). The "sleeping beauty" model (Ke et al., 2015) captures this pattern and suggests that a scientist's eureka moment may take decades to be validated and appreciated. A similar pattern observed in technology is named after "J-curve" theory, which suggests that revolutionary, general-purpose technologies (GPTs) like steam engines, electricity, and AI always take a long time to diffuse as they demand and grow complementary technologies, but once a supportive environment has bloomed, they dramatically extend productivity (Brynjolfsson et al., 2018). However, neither of these theories link the character of scientific discovery or technological innovation to their influential outcomes or confirm that novelty succeeds only over time. Our study identifies and recovers this missing piece of puzzle in the science of innovation.

The delayed recognition of radical innovations may inspire some to wonder whether it is possible to formulate science policies that accelerate the exploration, diffusion and application of transformative scientific ideas. We call for additional research fixed on exploring signals derived from publication data and metadata, at the level of individual papers, fields, and science as a whole, to empower institutional leaders, policy makers, and researchers within the Science and Information Metrics communities for use to direct and accelerate science (Hicks et al., 2015). The number of citations as a metric is short-sighted—it emphasizes short-term impact and not long-term influence (Wang et al., 2013). Novelty and disruption direct our gaze to the long-term impact of science, and our reformulation of novelty as the pointwise mutual information (PMI) of embedded journal vectors enable us to analyze the evolution of perceived novelty for the first time. We argue that the design, verification, and implementation of metrics that enable us to quantify and value novel failures on the long path to transformational success, will reduce the tyranny of short-term rewards that have unintentionally inspired narrow, incremental, redundant research.

## Acknowledgements

The authors thank members of the University of Chicago's Knowledge Lab and the Swarma Club (Beijing) for helpful discussions. We are grateful for support from AFOSR (#FA9550-19-1-0354), NSF (#1829366 and #1800956), DARPA (#HR00111820006), and Richard King Mellon Foundation.

# Supplementary Materials for

# New Directions in Science Emerge from Disconnection and Discord

Yiling Lin, James Evans, Lingfei Wu

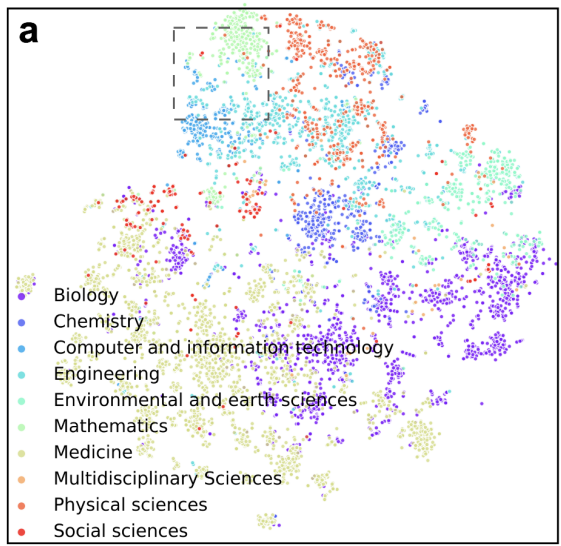 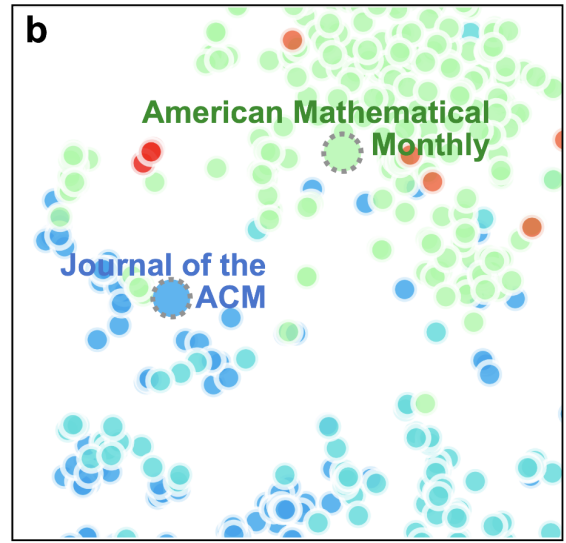

**Figure S1. Inner products in a journal embedding space are a proxy for z-score novelty.** (a) A journal embedding space constructed using the 2000 publications dataset (see methods for details), with journals (dots) colored by fields (using a different color scheme from Fig. 1). (b) A zoomed-in view of (a) highlighting Mathematics (green dots) and Computer Science journals (blue dots).

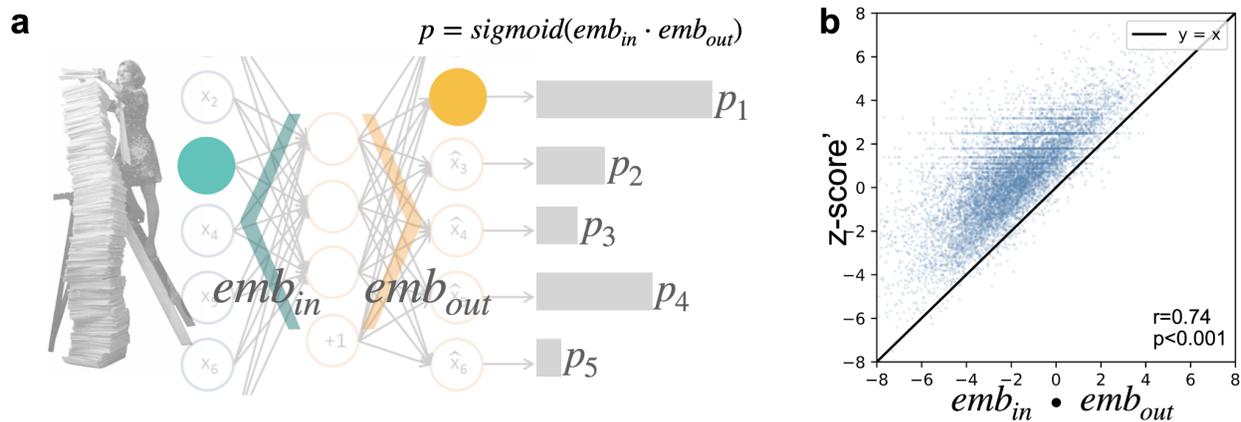

**Figure S2. Inner products in a journal embedding space are a proxy for z-score novelty.** (a) Illustration of prediction made a word2vec skip-gram model with negative sampling after the training process. Given a center journal (green), the trained model returns the probability $p$ of its context over the whole vocabulary. $p$ is modeled as the inner product between the in-embedding of the center journal and the out-embedding of the context journal over the sigmoid activation function. (b) Relationship between symmetric inner dot product and the revised z-score (see Eq. 3). The Pearson correlation coefficient is 0.74 and the p-value is much smaller than 0.001.

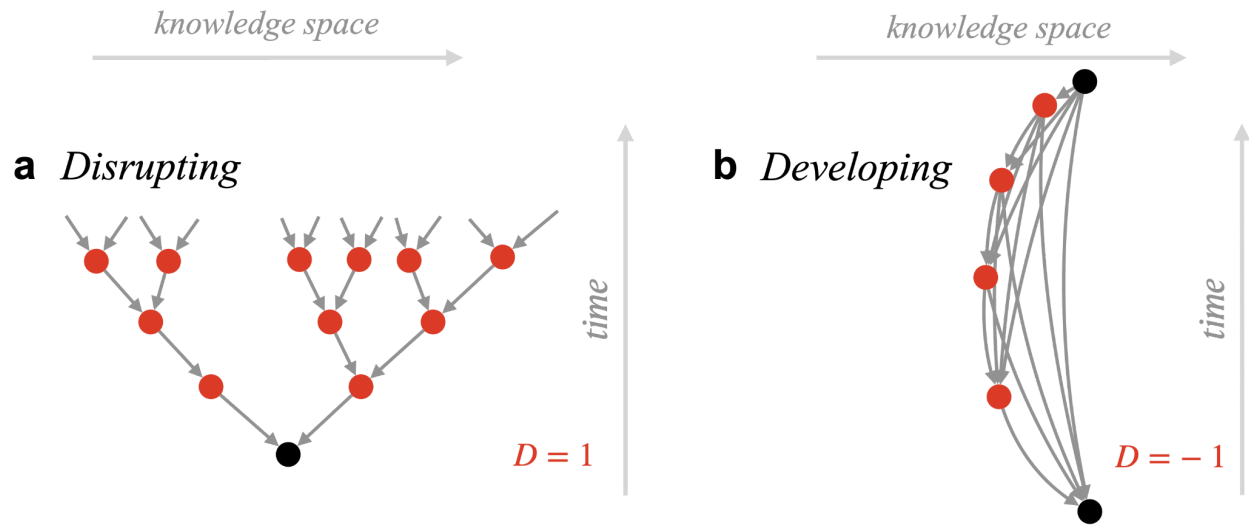

**Figure S3. Illustration of two extreme citation structures of science.** Science expands if the majority of papers (red circles) are disruptive (D = 1, Panel a) and collapses if developing (D = -1, Panel b). In the first scenario, science keeps branching out to cover new topics, but no consensus can be achieved. In the second scenario, science keeps returning back to the same topic and new papers are not significantly different from old papers after many generations—consensus is well established, but no new ideas are possible.

**Table S1. The asymmetry between disruptive and developing citations conditioned by paper novelty holds across disciplines.** For the 87,475 papers published in 1970, 45,677 (52%) among them are matched with fields of study provided by Microsoft Academic Graph. We select 11 fields of 100 or more papers. For each field we select the most novel (bottom 10 z-score) and conventional papers (top 10 z-score). For each paper in these two groups, we calculate the fraction of disruptive citations in the last year of analysis (2018). We then apply a statistical test (t-test) on the difference between the means of the average disruptive-citation fraction between the two groups.

| Field | N of papers | N of novel papers (bottom 10% z-score ) | N of conventional papers (top 10% z-score ) | Average of disruptive-citation fraction at the paper level in the novel group | Average of disruptive-citation fraction at the paper level in the conventional group | The difference between novel and conventional papers in the fraction of disruptive citations | T-test of the difference |
|---|---|---|---|---|---|---|---|
| Biology | 12587 | 1259 | 1240 | 0.508 | 0.395 | 0.113 | 10.09*** |
| Chemistry | 12052 | 1205 | 1204 | 0.519 | 0.512 | 0.007 | 0.62 |
| Medicine | 6253 | 624 | 627 | 0.554 | 0.454 | 0.100 | 5.99*** |
| Physics | 5298 | 529 | 313 | 0.492 | 0.398 | 0.094 | 4.39*** |
| Psychology | 3139 | 314 | 331 | 0.496 | 0.336 | 0.160 | 6.51*** |
| Mathematics | 2491 | 247 | 249 | 0.540 | 0.430 | 0.110 | 4.00*** |
| Materials Science | 1435 | 144 | 142 | 0.549 | 0.469 | 0.080 | 2.29* |
| Geology | 986 | 99 | 97 | 0.547 | 0.429 | 0.118 | 2.87** |
| Computer Science | 398 | 40 | 34 | 0.614 | 0.399 | 0.215 | 3.07** |
| Engineering | 273 | 27 | 31 | 0.654 | 0.356 | 0.298 | 4.44*** |
| Economics | 270 | 27 | 26 | 0.722 | 0.373 | 0.349 | 3.95*** |
| Sociology | 267 | 27 | 26 | 0.750 | 0.541 | 0.209 | 2.66* |

* p-value < 0.05; ** p-value < 0.01; *** p-value < 0.001.